# Nanoscale Manipulation of Single-Molecule Conformational Transition through Vibrational Excitation

Weike Quan[1,2,‡], Zihao Wang[1,2,‡], Yueqing Shi[1,‡], Kangkai Liang[1,2], Liya Bi[1,2], Hao Zhou[1], Zhiyuan Yin[1], Wan-Lu Li[2,3], Shaowei Li[1,2,*]

[1] Department of Chemistry and Biochemistry, University of California, San Diego, La Jolla, CA 92093-0309, USA

[2] Program in Materials Science and Engineering, University of California, San Diego, La Jolla, CA 92093-0418, USA

[3] Aiiso Yufeng Li Family Department of Chemical and Nano Engineering, University of California, San Diego, La Jolla, CA 92093-0448, USA

**ABSTRACT:** On-demand control of molecular actions is essential for realizing single-molecule functional devices. Such a control can be achieved by manipulating interactions between individual molecules and their nanoscale environment. In this study, we manipulate the conformational transition of a single pyrrolidine molecule on a Cu(100) surface by exciting its vibrations with tunneling electrons using scanning tunneling microscopy. Multiple transition pathways between two structural states are identified to be driven by distinct vibrational modes, whose corresponding nuclear motions are determined by density functional theory calculations. Tip-induced van der Waals forces and intermolecular interactions enable precise tuning of molecule-environment interactions, allowing modulation of vibrational energies, alteration of transition probabilities, and selection of the lowest energy transition pathway. This work reveals how external force fields in a tunable nanocavity can modulate molecular conformational transitions, offering an approach to deliberately engineer molecule-environment interactions for specific molecular functions.

## INTRODUCTION

Molecular conformational transitions are critical in various physical, chemical, and biological processes[1-8]. The structural changes of macromolecules[9] govern many essential life processes[10] from adenosine triphosphate (ATP) synthesis[2] to protein formation[11]. Similarly, the tautomeric states of small organic molecules affect their interactions with others and chemical reactivity[3,12,13], impacting both fundamental science[14] and pharmaceutical industry[15,16]. On-demand control of molecular conformation is therefore essential for the precise manipulation of chemical processes[15]. These conformational states are largely determined by the molecule's interactions with the nanoscale chemical environment[17-21] which influences the intramolecular bonding[22,23], ground states configuration[24] and the excitation pathways[25]. Consequently, nano-scale manipulation of molecule-environment coupling offers a promising route for on-demand control of the molecular conformation.

Scanning tunneling microscopy (STM) offers atomic-scale controllability for probing and manipulation of individual molecules on solid surfaces[26-28]. The nano-cavity between the STM tip and a substrate can be fine-tuned with sub-angstrom precision, allowing on-demand modification of the local chemical environment of the single molecule in it[29,30]. Specifically, tip-induced van der Waals forces have been applied to control the chirality[31] and adsorption geometry[32] of adsorbates. By combining microscopic structural imaging with the spectroscopical characterizations with scanning tunneling spectroscopy (STS) or inelastic electron tunneling spectroscopy (IETS), the effect of tip-molecule interaction on molecular electronic and vibrational properties can be precisely characterized[33]. This combination of in-situ tuning and molecular characterization allows the STM tip to deliberately manipulate the nanoscale environment around individual molecules.

STM-action spectroscopy (STM-AS) measures the action rate of a molecule within the STM cavity as a function of sample bias, providing energy-resolved insights into reactions of single molecules on surfaces[34,35]. Physical actions like hopping[36], rotation[27,37], structural switching[38,39] and chemical reactions like dissociation[40], isomerization[41], and hydrogen activation[42] can be triggered by exciting vibrational or electronic states with inelastic tunneling electron[34] or photon[41,42]. These excitations result in geometry changes, detectable as current variations in STM[36-38]. By plotting the action rate against stimulus energy, we can identify the energy thresholds corresponding to the vibration or electronic excitations that drive these actions or reactions[43-45].

Here, we applied the STM-AS to investigate the conformational transition of a single pyrrolidine molecule, a five-membered nitrogen heterocycle ring molecule of a vital role in biosystem and drug discovery[46], adsorbed on a Cu(100). Transition pathways between two structural states are driven by distinct molecular vibrations, identified through comparison with density functional theory (DFT) simulations. To control these transitions, we modulated the chemical environment within the STM cavity by adjusting the van der Waals forces exerted by the tip on the molecule, which in turn altered the potential energy surfaces and the vibrational mode energies responsible for mediating these transitions. Furthermore, attaching another pyrrolidine molecule to the STM tip introduced intermolecular steric repulsion, which altered the transition behavior and created a new transition pathway when the two molecules were brought into proximity.

RESULTS AND DISCUSSION

Individual pyrrolidines on Cu (100) surface exhibit bistable adsorption conformations designated as "high" (H) and "low" (L) states, easily distinguishable in STM images with characteristic heights of ~1.5 Å and 0.85 Å, respectively (Figure 1A). These states correspond to two structural energy minima, with the L state being 23.6 meV higher than the H state (Figure 1B) according to DFT simulation. This energy difference is significantly larger than the reported value of gas-phase pyrrolidine[47], primarily due to the interaction with the Cu substrate. Under the experimental temperature of 5.5 K, transitions between these two states cannot be spontaneously triggered by thermal excitation. However, they can be reversibly induced by scattering with inelastic tunneling electrons [48] or photons [49, 50], detected as distinct changes in the STM tunneling current as shown in Figure 1F.

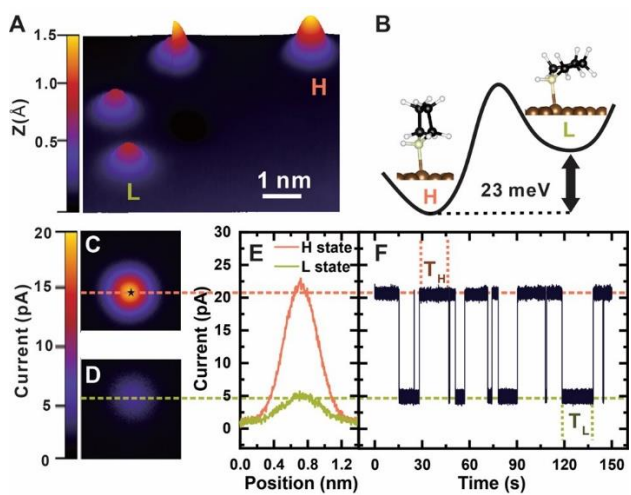

Figure 1. Structural transitions between two configurations of single pyrrolidine molecules adsorbed on Cu(100). (A). Constant current topographic image showing two configurations of pyrrolidine adsorbed on Cu(100). The labels 'L' and 'H' indicate the two molecules in "Low" and "High" states, respectively. Scanning conditions for the images were 50 mV and 20 pA. (B). Schematic representation of the double-well potential landscape illustrating the two adsorption configurations, with a 23 meV energy difference. (C-D). Constant height images of single pyrrolidine molecules in the H (C) and L (D) states. Scale: 1.3Å × 1.3 Å. (E). Line-cut profiles across the topographic images in (C) and (D), along the orange and green dashed lines, respectively. (F). Current time trace showing the switching of the molecule between the two states, with the tip positioned over the molecule at the location marked by a star in (C). The labels $T_H$ and $T_L$ represent the residence times of the molecule in the H and L states, respectively.

We investigated the molecule's switching between these two configurations by monitoring the current changes at a selected bias in the constant height mode to minimize influences from tip-substate distance variations. Figures 1C-D show topographic images of pyrrolidine in the H and L states, respectively. The highly symmetric images (Figures 1C-D) and line-cut profiles (Figure 1E) indicate a continuous in-plane rotation of the adsorbed molecule between the four equivalent adsorption geometries in both states due to the 4-fold symmetry of the Cu(100). The noisier L-state profile suggests a larger rotational energy barrier, leading to a slower and detectable rotation (see **Supporting Information** Section 3.1). The residence time at a certain bias for each state before switching to the other is analyzed through the current-time trace (Figure 1F).

The transition rate between two states increases with incremental electron energy, while exhibiting a steeper slope as new transition pathways open[43, 51]. By measuring the transition rate as a function of sample bias, six step-like thresholds are observed between 30 and 130 mV (indicated by arrows in Figure 2A). The energies of these thresholds were quantified using conventional fitting methods, as detailed in the **Supporting Information** Section 1.2[43, 51]. The H-to-L and L-to-H transitions account for 4 and 2 of the observed thresholds, respectively, based on analysis of the bias dependence of the residence times in each state separately (Figure 2B). Each threshold corresponds to the activation of a new transition pathway, associated with the excitation of a molecular vibration. It is important to note that these thresholds do not account for all possible transition pathways within this energy range. They are specifically related to excitation pathways that can be more easily triggered by inelastic tunneling electrons.

To reveal the nature of these molecular vibrations, DFT calculations were employed to simulate the vibrational modes of the H and L states, as shown in Figures 2E-J. The lowest energy transition-associated vibrational modes in either direction involve significant motion of the nitrogen atom (Figures 2E and G). In contrast, the higher energy modes are more associated with the motion of carbon atoms on the ring (Figures 2F and 2H-J). The spatial distribution of the measured transition yield, primarily determined by the excitation cross-section of the electron-scattered molecular vibration, varies depending on the dominating transition pathways. With the bias set to below 70 mV, the transitions in both directions are dominated by the two lowest energy modes primarily concerning the motion of the nitrogen atom (Figures 2E and 2G). The maximum yield of the L-to-H transition is concentrated at the center of the molecule, coinciding with the nitrogen adsorption site (Figure 2C). At 90 mV, where the transitions are dominantly mediated by the ring modes, the maximum yield of L-to-H transition shifts away from the molecular center (Figure 2D), agreeing with the location of the outer carbon atoms in the L state.

The variation of transition preference through distinct pathways at different bias provides a tunable mechanism for manipulating the preferred molecular conformation, as reflected by the bias dependence of the state occupation ratio ($Ratio_L = \frac{T_L}{T_L + T_H}$.). As the bias increases, the yield of the H-to-L transition begins to rise beyond the 40 mV bias threshold where the first H-to-L transition pathway becomes active. This results in an increasing L state occupation ratio until 58 mV, at which point the first L-to-H transition pathway is activated. The competition between these two pathways leads to a maximum L state occupancy near 60 mV bias (Figures 2K-L).

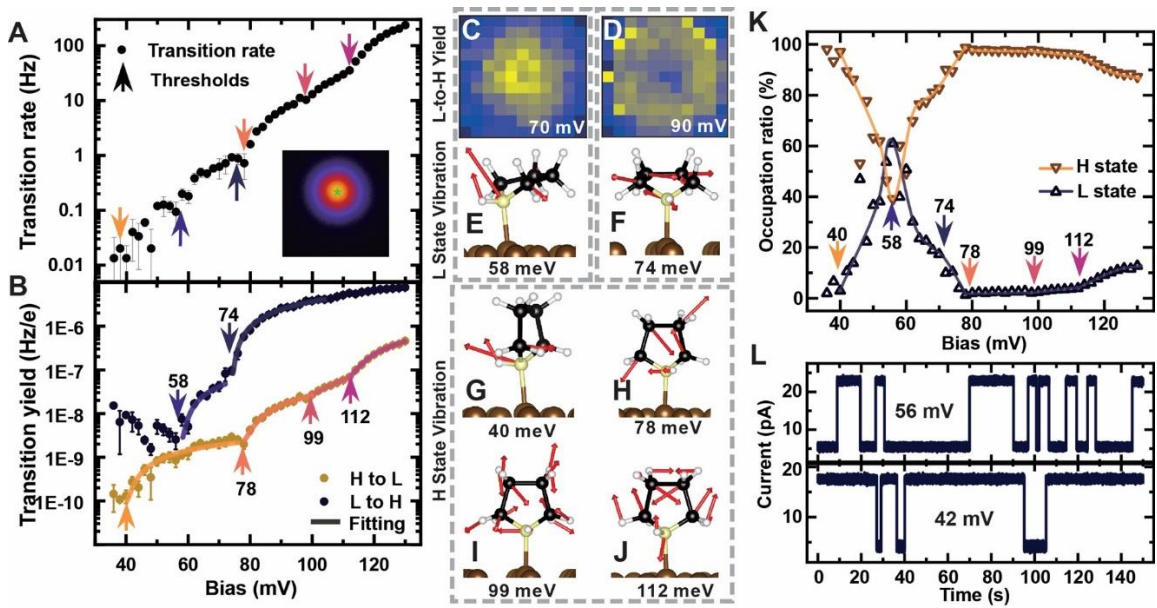

**Figure 2.** Dependence of the structural transition rate on the energy of tunneling electrons. (A). Total transition rate, including both H-to-L and L-to-H transitions, as a function of electron energy ranging from 30 to 130 meV, showing six step-like thresholds indicated by arrows. (B). Normalized transition yields for the H-to-L and L-to-H transitions, respectively. (C-D). Spatially resolved mapping of the L-to-H transition at bias voltages of 70 mV and 90 mV, respectively. Mapping scale: 9 Å × 9 Å. (E-F). DFT simulations of the two vibrational modes at 58 meV (E) and 74 meV (F) for the L state, corresponding to the two transition thresholds in (B). (G-J). Four simulated vibrational modes of the H state related to the four thresholds of the H-to-L transition yield in (B) at energies of 40, 78, 99, and 112 meV, respectively. (K). Occupation ratio of H and L states as a function of sample bias. (L). Current trace at different bias, taken at the same position at a constant height. The molecule favors the H state at 42 mV bias but prefers the L state at 56 mV.

The molecular conformation can be further manipulated by adjusting the local chemical environment in the STM cavity through changes in the tip-molecule separation. The STM tip exerts an attractive van der Walls force on the molecule[31-33], which, in this case, stabilizes the H state while destabilizing the L state, resulting in a more asymmetric potential profile (Figure 3A). This effect is observed as a reduction of the L state residence time and the increase in the H state residency at the same bias (Figures 3C and G). Additionally, the overall transition yield (Figure 3B) is also affected by the change in the energy barrier between the L and H states. The H-to-L transition yield decreased by three orders of magnitude as the tip approached by ~1.6 Å, while the L-to-H transition yield increased by approximately one order of magnitude.

The tip-induced force affects each molecular vibration in distinct ways, leading to different variations in the corresponding transition pathways. This tip-molecule interaction is closely tied to the molecular dipole, particularly in relation to the bond involved in the corresponding vibration[33]. For the two L-to-H transition pathways corresponding to the L state vibrations (Figure 3E), the threshold of the lower energy pathway (~60 meV, related to the large nitrogen motion) significantly redshifts, while the higher threshold (~74 meV, featured as the ring motion) remains unaffected with the tip approached about 1.4 Å. Conversely, for the transition from H to L states, the threshold corresponding to large nitrogen motion (~40 meV) undergoes a significant blue shift, while the vibrations at 78 meV and 110 meV exhibit minor redshifts as the tip approaches (Figure 3D). These differential effects on vibrational modes are related to the molecular dipole distribution and the atomic motions involved in the vibrations. For pyrrolidine on Cu(100), the dipole moment is concentrated around the nitrogen atom, particularly involving polar bonds involving C-N and N-metal. Since the N-metal bond position remains unchanged during the conformational change (Figure 3A), the tip-induced van der Waals force primarily influences the nitrogen atom, leading to the significant shift of the two lowest energy vibrational modes concerning the large motion of the C-N bond as depicted in Figures 2E and 2G. The higher energy modes, however, are mainly associated with the motion of carbon atoms, and are therefore less sensitive to the tip-molecule interaction.

The direction of nitrogen atom motion in the L and H states determines whether the lowest-energy vibrational mode of the corresponding state undergoes red or blue shifts as the tip approaches. In the H state, this vibration involves a larger in-plane motion (Figure 2G), while in the L state features a larger out-of-plane motion (Figure 2E). Consequently, the attractive force from the tip weakens the N-metal bond, decreasing the energy of vibrations with significant out-of-plane motion, while increasing the energy of vibrations with substantial in-plane motion. Similar effects have been observed in other molecules with comparable dipole moments [33]. Additionally, due to the shift of the molecular vibration, the bias of where the maximum L state occupancy decreases as the tip approaches the molecule (Figure 3F).

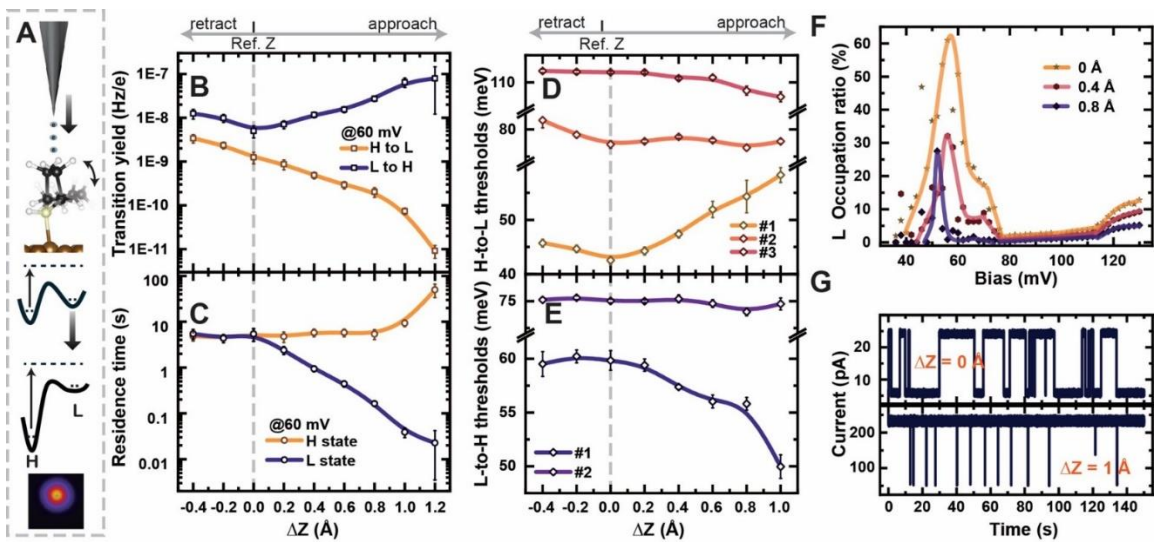

**Figure 3.** Modification of threshold energies for different transition pathways due to tip-molecule interactions. (A). Schematic illustrating how tip-molecule interactions modify the potential landscape for a molecule residing in the STM cavity. As the tip approaches the molecule, the double-well potential becomes increasingly asymmetric. (B). Transition yields for H-to-L and L-to-H transitions as a function of tip-substrate distance, with Δz varying from 0.4 Å retraction to 1.0 Å approach relative to the z reference point (ref. z) set by 50 mV and 20 pA, with a bias maintained at 60 mV. (C). Average residence times of the H and L states at different tip-substrate distances, measured under the same conditions as (B). (D-E). Threshold energies for H-to-L and L-to-H transitions as a function of tip-molecule separation. (F). Occupation ratio of the L state at varying tip-molecule separations. (G). The current traces taken at different Δz, with a bias maintained at 60 mV. The molecule favors the H state with the tip approached by 1 Å from the ref. z.

In addition to tip-molecule interaction, intermolecular interaction offers another means to manipulate the molecular conformational states. By picking up a pyrrolidine molecule with the STM tip and moving it to the top of another molecule on the surface, the intermolecular interaction can be deliberately controlled by changing the tip-substrate distance. The pyrrolidine adsorbed on the tip apex also exhibits two configurations and is found to interconvert more readily. Notably, the presence of a tip-adsorbed molecule significantly increases the transition barrier of the substrate molecule, making it difficult to switch states at biases below 80 mV. Additional discussion regarding this effect is included in the **Supporting Information** Sections 3.5-3.7. Here, we focus on the switching behavior of the tip-adsorbed molecule within our detection range, which shows the lowest energy transition thresholds of ~40 meV for H-to-L transitions and ~52 meV for L-to-H transitions.

As illustrated in Figure 4A, intermolecular interaction involves different mechanisms at different intermolecular distances. At a relatively larger distance, the intermolecular interaction is primarily driven by dipole-dipole interaction, which manifests as a long-range attractive force. As two molecules get closer to each other, the saturated carbon rings induce a strong steric repulsion between the two molecules. We measured molecular transitions by varying the tip distance by -0.2 to 1.0 Å relative to a reference setpoint of 50 mV and 20 pA as shown in Figures 4B and 4C. From -0.2 to 0.4 Å, the attractive force increases, extending the H state lifetime and raising the H-to-L transition barri-er. Beyond 0.4 Å, short-range repulsion impacts the H state, reversing transition yield and lifetime trends and altering the H state structure (see **Supporting Information** Section 3.6). Conversely, the L state of the tip molecule in this range continues to experience increasing attractive forces as the steric pressure from the carbon ring is relieved when it moves away from the other molecule. Upon approaching the tip by approximately 1 Å, the steric pressure between the two molecules eventually prevents the L-to-H switching of the tip-bound molecule when the surface molecule is in the H state. In this scenario, the tip-bound molecule can only switch from L to H after the surface molecule transitions to the L state, which requires a higher energy. Additional details on this process are provided in the **Supporting Information** Section 3.7.

Intermolecular interactions also modify molecular vibrations, resulting in the change of the most energetically favorable transition pathway. Similar to a bare tip, the lower energy mode concerning large nitrogen motion is sensitive to the intermolecular distance. For the H-to-L transitions, within the range of -0.2 to 0.4 Å where attractive forces dominate, the lowest energy threshold exhibits a blueshift, similar to that observed with the W tip. However, at distances of 0.4 to 0.9 Å, where repulsive forces become significant, this trend reverses, resulting in an energy redshift. Additionally, a new, lower energy threshold emerges as an additional transition channel opens (Figure 4D). This could result from changes to the potential energy surface of the H state due to the repulsive intermolecular forces, which raise the energy of the H state and create a more symmetric double-well potential. The lowered energy barrier makes it possible to mediate the H-to-L transition with an even lower energy vibration, which becomes the new, most energy-preferable transition pathway (Figure 4A). For switching from the L state, where the long-range attractive force is predominant throughout the entire range, the vibration energy consistently shows a red shift as the tip approaches (Figure 4E). We note that the detailed features in the threshold energy shifts in response to the tip-substrate distance—such as the peak at Δz = 0.3 Å and the dip at Δz = 0.6 Å in Figure 4E—vary depending on the tip conditions used during measurements. These variations are likely linked to the specific profile of the potential energy landscape, which is influenced by the tip geometry. However, the overall trends remain consistent across all measurements, regardless of tip conditions: a decrease in the L-to-H threshold energy and an initial increase followed by a decrease in the H-to-L threshold energy as the distance between two molecules decreases.

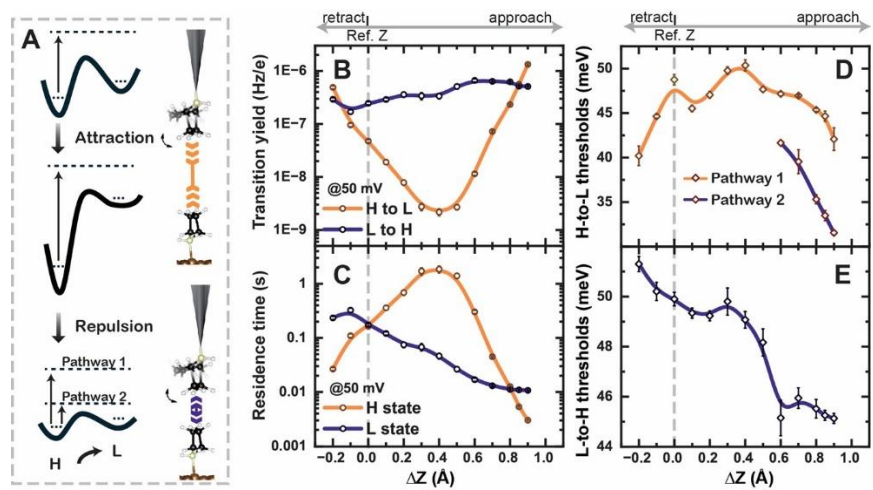

**Figure 4.** Alteration of molecular structural transitions by intermolecular interactions. (A). Schematic of the STM junction, depicting one pyrrolidine molecule adsorbed on the tip and another on the surface. The two molecules attract each other at greater distances and repel each other when brought closer. (B). H-to-L and L-to-H transition yields of the molecule on the tip recorded at a bias of 50 mV, measured at various tip-substrate separations. (C). Average residence times of the H and L states for the tip molecule at a 50-mV bias across different tip-substrate separations. (D-E). Threshold energies for L-to-H and H-to-L transitions of the tip molecule at varying tip-substrate separations. The lowest energy transition pathway switches from the curve indicated by orange to the one indicated by purple when the tip approaches 0.6 Å relative to the ref. z.

## Conclusion

In conclusion, this study elucidates the vibration-mediated pathways governing the conformational transitions of a bistable pyrrolidine molecule adsorbed on a Cu(100) surface. By utilizing the tunable nanocavity formed between the STM tip and the metal substrate, we have demonstrated how sample bias, tip-molecule van der Waals forces, and intermolecular interactions can precisely control the transition energy barriers and the molecular vibrations that govern the transition pathways between the molecule's two states. Our findings offer valuable insights into how molecular transitions can be manipulated through fundamental interactions including van der Waals attraction and steric repulsion, shedding light on the on-demand control of molecular reaction dynamics within a tunable nano-environment. The ability to harness these external forces to control molecular switches paves the way for designing functional single-molecule devices with the desired functionality.

## ASSOCIATED CONTENT

**Supporting Information**. The supporting information is available free of charge.
Details of experimental and computational methods and additional data (Figure S1 to S9)

## AUTHOR INFORMATION

### Corresponding Author

Shaowei Li – Department of Chemistry and Biochemistry, University of California, San Diego, California 92093-0309, United States; Program in Materials Science and Engineering, University of California, San Diego, California 92093-0418, United States;
Email: shaoweili@ucsd.edu

### Author Contributions

‡These authors contributed equally.
### Notes
The authors declare no competing financial interest.

## ACKNOWLEDGMENT

This work was supported by the United States National Science Foundation (NSF) under Grant No. CHE-2303936 (to Shaowei Li). The authors acknowledge the use of facilities and instrumentation supported by NSF through the UC San Diego Materials Research Science and Engineering Center (UCSD MRSEC) with Grant No. DMR-2011924. This work used the San Diego Supercomputer Center (SDSC) Expanse at UC San Diego for theoretical calculations through allocation CHE-240050 from the Advanced Cyberinfrastructure Coordination Ecosystem: Services & Support (ACCESS) program, which is supported by NSF Grants No. OAC-2138259, No. OAC-2138286, No. OAC-2138307, No. OAC-2137603, and No. OAC-2138296.

SYNOPSIS TOC

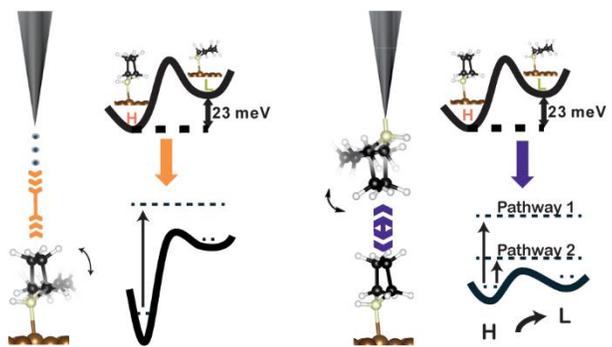

For Table of Contents Only